\documentclass[11pt]{article}
\usepackage{graphics,epsfig}
\usepackage[latin1]{inputenc}
\usepackage[english,activeacute]{babel}
\usepackage[]{graphicx}
\usepackage{latexsym}
\usepackage{amsmath, amsthm, amsfonts, amssymb}
\usepackage{verbatim}
\usepackage{color}

\begin{document}

\newtheorem{theo}{Theorem}[section]
\newtheorem{definition}[theo]{Definition}
\newtheorem{lem}[theo]{Lemma}
\newtheorem{prop}[theo]{Proposition}
\newtheorem{coro}[theo]{Corollary}
\newtheorem{exam}[theo]{Example}
\newtheorem{rema}[theo]{Remark}
\newtheorem{example}[theo]{Example}
\newtheorem{principle}[theo]{Principle}
\newcommand{\ninv}{\mathord{\sim}}
\newtheorem{axiom}[theo]{Axiom}

\title{Natural information measures in Cox' approach for contextual probabilistic theories}

\author{{\sc Federico Holik}$^{1}$ \ {\sc ,}\ {\sc Angel Plastino}$^{1,\,3}$ \ {\sc
 and} {\sc Manuel S\'{a}enz}$^{2}$}

\maketitle

\begin{center}

\begin{small}
1- Universidad Nacional de La Plata, Instituto
de F\'{\i}sica (IFLP-CCT-CONICET), C.C. 727, 1900 La Plata, Argentina \\
2- Department of Mathematics, University of Buenos Aires \& CONICET. \\
3- Universitat de les Illes Balears and IFISC-CSIC, 07122 Palma de
Mallorca, Spain.
\end{small}
\end{center}

\vspace*{0.21truein}

\abstract{In this article we provide, from a novel perspective, arguments
that support the idea that, in the wake of Cox' approach to
probability theory, von Neumann's entropy should be the natural
one in Quantum Mechanics. We also generalize the pertinent
reasoning to more general orthomodular lattices, which reveals the
structure of a general non-Boolean information theory.}

\vspace*{10pt}

\begin{small}
\centerline{\em Key words: von Neumann entropy, Information Theory, Lattice Theory, Non-Boolean Algebras}
\end{small}

\section{Introduction}
The problem of characterizing information measures has puzzled
people since the very beginning of information theory. As an
example, this intriguing character is expressed in the very origin
of the term `entropy' for Shannon's measure. In Shannon's words:

\begin{quotation}
My greatest concern was what to call it. I thought of calling it an ``information'', but the word was overly
used, so I decided to call it an ``uncertainty''. When I discussed it
with John von Neumann, he had a better idea. Von Neumann told me,
``You should call it entropy, for two reasons. In the first place
your uncertainty function has been used in statistical mechanics
under that name, so it already has a name. In the second place, and
more important, nobody knows what entropy really is, so in a debate
you will always have an advantage''.\cite{Tribus71} (see also
\cite{Bengtsson2006}, page 35).
\end{quotation}

This entropic {\it mystery}  did nothing but grow since the advent
of quantum information theory (QIT) \cite{NielsenBook}, in which von Neumann's entropy (VNE) plays a significant role (as for example, in the \emph{quantum coding theorem} presented in
\cite{Schumacher-QuantumCoding-1995,Jozsa-Schumacher-NewProof-1994}).
Of course, von Neumann's measure is not `alone': there are many
other entropic measures which also play a significant role in QM and
QIT, such as the Tsallis' \cite{TsallisEntropy1988,RasteginQIC-2014} and Reny\`{i}'s
\cite{renyi1961}
entropies, and even more general ones
\cite{Rossignoli-2010,Zozor-Bosyk-Portesi-2014,MullerRenyi}. This
fact led to an important debate about which is the \emph{correct}
information measure for the quantal realm (see for example
\cite{Zeilinger-AgainstShannon-2001} and
\cite{Timpson-AgainstZeilinger}). Many studies attempted to
characterize Shannon's entropy first
\cite{Aczel-WhyShannon-1974,AczelBook-1966,Ochs-UniqueCharacterization-1975,Ochs-UniqueCharacterization-1976},
but also von Neumann's \cite{Ochs-VN-1975}. In this article, we
offer a novel perspective which considers VNE as the natural
information measure for a non-Boolean probability calculus.

Classical information theory (CIT) relies on the notion of
probability: as stressed by Shannon, the probabilistic aspect of
the source is at the basis of his seminal work \cite{Shannon1948}.
It is widely accepted that classical probability theory can be
axiomatized using Kolmogorov's postulates
\cite{KolmogorovProbability}. But it turns out that there exists
another approach to classical probability, namely, that of R. T.
Cox \cite{Cox-ProbabilityFrequency-1946,CoxLibro}. In Cox'
approach, probabilities are considered as an inference calculus on
a Boolean algebra of propositions: a rational agent, intending to
make inferences using classical logic (wherefrom the Boolean
structure emerges), must compute the plausibility of certain
events to occur. It turns out that the only measure of
plausibility compatible with the algebraic symmetries of the
Boolean algebra of propositions is
---up to rescaling--- equivalent to Kolmogorov's probability theory
\cite{Knuth-DerivingLaws,Goyal-Knuth-Symmetry,Knuth-Neurocomputing-2005}.
In this way, the plausibility calculus is considered as a direct
extension of classical deduction theory to an inference theory:
the extension of rationality applied to the calculus of
plausibility.

In his preliminary works, Cox also conjectured that  Shannon's
entropy \cite{Cox-ProbabilityFrequency-1946} was the natural
information measure for classical probability distributions. This
approach was considerably developed and improved in
\cite{Knuth-DerivingLaws,Goyal-Knuth-Symmetry,Knuth-Neurocomputing-2005,Knuth-MeasuringQuestions,knuth-aistats2005,KnuthWhatIsAQuestion,Knuth-Skilling-Axioms-2012}.
In this way, Shannon's and Hartley's entropies have been
characterized as the only entropies that can be used for the
purposes of inquiry, in the sense that other entropies will lead
to inconsistencies with the Boolean character of the lattice of
assertions \cite{knuth-aistats2005}.

In \cite{Holik-Plastino-Saenz-AOP-2014}, we presented a derivation
of the axioms of non-commutative probabilities in Quantum Mechanics
(QM) by appealing to the non-distributive (non-Boolean) character of
the lattice of projections of the Hilbert space (Cf. Appendix
A of this work for elementary notions of lattice
theory). This was done by extending  Cox' approach for the
orthomodular lattice of projection operators to i) the quantal case
and ii) more general non-Boolean algebras. In {\it this work} we
complement the approach by deriving the VNE as the most natural
information measure in the quantum context. As in
\cite{Holik-Plastino-Saenz-AOP-2014}, we extend our results to more
general (atomic) orthomodular lattices. This is done by exploiting
the fact that Cox' derivation of Shannon's entropy can be applied to
all possible maximal Boolean subalgebras of an arbitrary atomistic
orthomodular lattice. Thus, according to our extension of Cox'
approach to the non-commutative realm, the VNE and Measurement
Entropy (ME) \cite{Short-Wehner-2010,Barnum2010,HeinEntropyInQL1979}
arise as the most natural information measures.

The results presented in this work pave the way for a new way of
conceiving information theory. While classical probabilities give
rise to Shannon's theory, and thus, lead to CIT, non-Boolean
probabilities give rise to VNE and QIT. Thus, QIT could be
conceived as a non-Boolean generalization of CIT. This opens the
door to a new way of exploring physical theories from the
informational point of view, due to the fact that probabilistic
theories more general than the ones appearing in classical and
(standard) quantum mechanics can be conceived.

Indeed, during the 30's von Neumann developed a theory of rings of
operators \cite{vN-OriginalPaperOnVNAlgebras-1930} (today known as
von Neumann algebras
\cite{RedeiSummersQuantumProbability,RedeiQuantumLogicInAlgebraicApproach}),
and subsequently, in a joint work with Murray, they provided a
classification of factors
\cite{Murray-VN-OnRingsOfOperatorsI,Murray-VN-OnRingsOfOperatorsII,VN-OnRingsOfOperatorsIII,Murray-VN-OnRingsOfOperatorsIV}.
While quantum systems of finite degrees of freedom (as is the case
for example, in standard non-relativistic QM), can be described
using Type I factors, more general Factors are needed for more
general theories: it can be shown that Type III Factors must be used
in relativistic quantum field theory, and Type II factors may appear
in quantum statistical mechanics of infinite systems
\cite{Yngvason2005-TypeIIIFactors,RedeiSummersQuantumProbability}.
Thus, it is expected that theories to be yet developed, such as a
quantum theory of gravity, may very likely imply the use of more
general probabilistic models (perhaps not contained in the above
examples).

The more general\footnote{For the case of negative probabilities, see for example
 \cite{AcacioNegative,AcacioNegative2}.} framework for studying probabilistic models up to
now is provided by the Convex Operational Models (COM) approach
\cite{Holik-Plastino-ReviewNOVA-2014,Holik-Plastino-Saenz-ReviewHindawi}.
In a general probabilistic model, the probabilities will not be
necessarily Kolmogorovian (as is the case for the probabilities
appearing in Type I, II and III factors). Thus, we envisage the
development of a \emph{non-Kolmogorovian (or generalized)
information theory}. A clear example of the fact that such an entity
does exists can be found in studies focusing on the validity of
informational notions such as
\cite{Short-Wehner-2010,Barnum-CloningAndNoBroadcasting,Barnum2010,HolevoBook-2011}
(cf. \cite{Holik-Plastino-Saenz-ReviewHindawi} for more references
on the subject). In this work, we show that CIT and QIT are just
particular cases of this approach; the first would be the Boolean
case, and the second, the one represented by the Type I factors of
the Murray-von Neumann classification theory, i.e., as the algebras
of bounded operators acting on separable Hilbert spaces.  In this
context, it is pertinent to mention that quantum algorithms where
shown to exploit the non-Boolean character of the lattice of
projection operators in quantum mechanics \cite{BubQLandQI}. The
generalization of the Bayesian Cox' approach presented here (and in
\cite{Holik-Plastino-Saenz-AOP-2014}), provides a unifying formal
framework for dealing with possible physical theories. It also
provides a possible interpretation of VNE and ME as natural measures
of information for non-commutative event structures. In other words,
as a natural information measure for theories exhibiting a highly
contextual character, like standard quantum mechanics
\cite{GuhneCabello-2010,CabelloProposalFor-2010}.

Before concluding this Introduction, we remark that an important advantage of extending the Cox' approach to non-Boolean settings is that it offers a novel argument in favor of the use of the logarithmic functional form appearing in the VNE and the ME. As we have remarked above, there is a debate around the question of why using the VNE instead of more general quantum information measures (such as the quantum versions of R\'{e}nyi and Tsallis entropies) in the quantum realm. Moreover, while the ME was introduced in References \cite{Short-Wehner-2010}, \cite{Barnum2010} and \cite{HeinEntropyInQL1979}, no conceptual argument is presented there in favor of using that functional form instead of more general ones (apart from observing that it possesses some of the `desired properties' shared by Shannon's measure and the VNE). In fact, in Section $4.3$ of reference \cite{Short-Wehner-2010}, a R\'{e}nyi functional variant of the ME is considered as another possible alternative for the purpose of studying quantum key encryption protocols. In other words, previous approaches do not focus in giving reasons for singularizing the VNE and the ME amongst other possibly useful choices. In this paper we show that in the framework of a rational agent looking for a measure of the questions unanswered by a particular probability assignment in a contextual probabilistic model, the VNE and more generally, the ME, appear as the most rational choices. From this `\textit{contextual rational agent}' perspective, the functional forms appearing in the VNE and the ME can be considered as the most reasonable choices compatible with the algebraic structure of a contextual inquiry calculus. Notice that this approach also allows for a very intuitive interpretation of the VNE and the ME, which was not present in previous works.

The paper is organized as follows. In Section
$2$ we start by
reviewing classical probability theory (in the approaches of
Kolmogorov and Cox) and probabilities appearing in QM, emphasizing
the differences with the classical case. In Section
$3$ we present a
digression on Cox' \cite{CoxLibro} and more recent \cite{Knuth-Neurocomputing-2005} derivations of Shannon's entropy as natural information measures for Boolean algebras. Next, in Section
$4$, we show how VNE arises as a
natural measure of information for the Hilbertian projection
lattice. In Section $5$, we
discuss generalized probabilistic models and ME. Finally, in Section
$6$ we draw some conclusions. Elementary notions
of lattice theory can be found in Appendix A.

\section{Axioms For Probability
Theory}\label{s:ClassicalProbabilityAndQuantumProbability}
\subsection{Kolmogorov's axiomatization of classical probability}
\noindent Let $\Sigma$ represent a sigma-algebra of subsets of a
given outcome set. To fix ideas, consider the example of a dice. For
this case, the outcome set $\Omega=\{1,2,3,4,5,6\}$ is the set of
all possible results, and $\Sigma=\mathcal{P}(\Omega)$ is the set of
all possible subsets of $\Omega$; each element of $\Sigma$
represents a possible event (for example, the event ``the result is
even", is represented by the set $\{2,4,6\}$ and so on).
Kolmogorov's axioms can be presented in the form of conditions on a
measure $\mu$ over $\Sigma$ as follows \cite{KolmogorovProbability}:

\begin{eqnarray}\label{e:kolmogorovian}
&\mu:\Sigma\rightarrow[0,1]&\nonumber\\
&\noindent \mbox{which satisfies}&\nonumber\\
&\mu(\emptyset)=0&\nonumber\\
&\mu(A^{c})=1-\mu(A),&\\
&\mbox{where}\,\,(\ldots)^{c}\,\,\mbox{stands for the set-theoretical complement.}\nonumber\\
&\mbox{For any pairwise disjoint denumerable family} \{A_{i}\}_{i\in I},\nonumber&\\
&\mu(\bigcup_{i\in I}A_{i})=\sum_{i}\mu(A_{i}).&\nonumber
\end{eqnarray}

With this minimal axiomatic basis the whole building of classical
probability theory can be erected.

A \emph{random variable} is defined as a function $X : \Sigma \longrightarrow \mathbb{R}$ that assigns real values to the elements of $\Sigma$. Random variables
are intended to describe properties of the system under study that
depend on the different possible outcomes that may result from a
given experiment. A random variable may be \emph{discrete} if its set of
possible values is countable, or \emph{continuous} if there exists a
continuous function which determines its probability distribution
according to

\begin{equation}
P( X \subseteq B ) = \int_{B} f(x) \ dx
\end{equation}

Although not necessarily, the formalization of probability given by
Kolmogorov's axioms is usually associated with an \emph{objectivist}
(of \emph{frequentist}) interpretation of probability theory, in
which probabilities represent a property of the system under study,
and are therefore capable of being subject to experimental test.

\subsection{Cox' approach to classical probability}\label{s:CoxProbability}

Alternatively, in Cox' approach probabilities are interpreted in a
subjective manner: they do not represent properties of physical
systems, but rather they are related to the information one
possesses about them. The aim of Cox was to establish probability
theory as a form of induction arising as an extension of classical
logic to situations of incomplete knowledge. As it will be shown,
by doing so, Cox arrives at the same results as the ones obtained
from Kolmogorov's axioms. However, these two approaches
significantly differ at the conceptual level. In this section,
although we will follow Cox' original deductions (presented in
\cite{Cox-ProbabilityFrequency-1946,CoxLibro}), for the sake of
clarity, we will somewhat change Cox' notation. In
\cite{CoxReview} and \cite{CoxCriticas}, a more detailed
discussion on Cox' work, together with implications and
criticisms, can be found.

\noindent Let us call $\mathbf{P}$ the set of propositions that a
rational agent uses to describe a system under study and ``$\neg$",
``$\vee$" and ``$\wedge$", the logical negation, disjunction, and
conjunction, respectively. Cox starts by  postulating the existence
of a function $\varphi_h : \mathbf{P} \longrightarrow \mathbb{R}$
that represents the \emph{plausibility} of the propositions in
$\mathbf{P}$ on the basis of a special knowledge possessed by the
agent. Such knowledge is that of a proposition, called  $h$ (usually
called \emph{hypothesis}), that i) happens to be true and ii)
satisfies:

\begin{itemize}
\item $\forall a \in \mathbf{P}, \ \varphi_h(\neg a) = f(\varphi_h(a))$, for some function $f : {\bf P} \longrightarrow \mathbb{R}$.

\item $\forall a, b \in \mathbf{P}, \ \varphi_h(a \vee b) = g[\varphi_h(a),\varphi_h(b)]$,
for some function $g: \mathbf{P} \times \mathbf{P} \longrightarrow \mathbb{R}$.
\end{itemize}

\noindent It is now possible to derive the calculus of probabilities
by imposing on this structure the symmetries of a Boolean algebra
\footnote{By \emph{classical logic} one refers to the propositional
calculus endowed with the operations ``$\neg$", ``$\vee$" and
``$\wedge$". It is widely known that the algebraic structures
corresponding to this propositional calculus are closely related to
Boolean algebras.}. On such a basis one arrives at results analogous
to the ones obtained from Kolmogorov's axioms.

By imposing coherence of the function $\varphi_h(\cdot)$ with the
associativity of conjunction ($a \wedge (b \wedge c) = (a \wedge
b) \wedge c)$), Cox showed that the function $g(x,y)$ must satisfy
the functional equation

\begin{equation}\label{eq:prod}
 g[x,g(y,z)] = g[g(x,y),z]
\end{equation}

Using the theory developed in \cite{AczelBook-1966}, it can be
shown that after a re-scaling and a proper definition of the
probability $P(a|h)$ in terms of $\varphi_h(a)$, this equation's
solutions lead to the \emph{product rule} of probability theory

\begin{equation}\label{eq:prod2}
 P(a \wedge b |h) = C P(a | h \wedge b) P(b|h)
\end{equation}

\noindent where $C$ is a constant. The definition of $P(a|h)$ in
terms of $\varphi(a|h)$ is omitted, as in actual computations one
ends up using only the function $P(a|h)$ and never $\varphi(a|h)$.
On the other hand, imposing coherence with i) the law of double
negation ($\neg \neg a=a$) and ii) Morgan's law for disjunction
($\neg (a \vee b) = \neg a \wedge \neg b$), Cox arrives to a
functional equation for $f(\cdot)$ which has solutions in terms of
$P(a|h)$ given (up to re-scaling) by

\begin{equation}\label{eq:neg}
 P(a|h)^r + P(\neg a|h)^r = 1
\end{equation}
\noindent This seemingly arbitrary choice of value for the
constant $r$ can be avoided via re-scaling probability to absorb
the $r$ exponent. That is to say, it can be avoided by defining
probability as $P'(a|h) \equiv P^r(a|h)$ instead of $P(a|h)$. Cox
decides to take $r=1$ and thus he obtains the usual rule for
computing the probabilities of complementary outcomes. Finally,
using results (\ref{eq:prod2}) and (\ref{eq:neg}), and imposing
coherence with i) the law of double negation and ii) Morgan's law
for conjunction ($\neg (a \wedge b) = \neg a \vee \neg b$), Cox
deduces the \emph{sum rule} of probability theory:

\begin{equation}
 \label{eq:sum}
 P(a \vee b|h) = P(a|h) + P(b|h) - P(a \wedge b|h)
\end{equation}
\noindent It can be easily shown from equations (\ref{eq:prod2})
and (\ref{eq:sum}) that, if normalized to $1$, $P(a|h)$ satisfies
all the properties of a Kolmogorovian probability (equations
\ref{e:kolmogorovian}).

\subsection{Axioms for probabilities in quantum
mechanics}\label{s:QuantumProbabilities}

In  \cite{feynman1951} R. P. Feynman defines probabilities as
follows:

\begin{quote}
I should say, that in spite of the implication of the title of this
talk the concept of probability is not altered in quantum mechanics.
When I say the probability of a certain outcome of an experiment is
$p$, I mean the conventional thing, that is, if the experiment one
expects that the fraction of those which give the outcome in
question is roughly $p$. I will not be at all concerned with
analyzing or defining this concept in more detail, for no departure
of the concept used in classical statistics is required.

What is changed, and changed radically, is the method of calculating probabilities.
\end{quote}

Feynman asserts that while the concept of probability is not altered
in QM, the method of calculating probabilities changes radically.
What does this mean? In order to clarify, let us write down things
in a more technical way. To begin with, a general state in QM can be
represented by a density operator, i.e., a trace class positive
hermitian operator of trace one
\cite{Holik-Plastino-PRA-2011,Holik-Zuberman-2013}. Let
$\mathcal{P}(\mathcal{H})$ be the orthomodular lattice of projection
operators of a Hilbert space $\mathcal{H}$ (cf. App. A).
Due to the spectral theorem, every \emph{physical event} (i.e., the
outcome of any conceivable experiment), can be represented as a
projection operator in $\mathcal{P}(\mathcal{H})$. If $P$ is a
projection representing an event and the state of the system is
represented by the density operator $\rho$, then, the probability
$p_{\rho}(P)$ that the event $P$ occurs is given by the formula

\begin{equation}
p_{\rho}(P)=\mbox{tr}(\rho P)
\end{equation}

\noindent which is known as Born's rule
\cite{Holik-Zuberman-2013}. Given an event $P$ and state $\rho$, if the experiment is repeated many times, Born's rule assigns a number which coincides with the fraction mentioned in Feynman's quotation. Gleason's theorem \cite{Gleason1975}
ensures that density operators are in bijective correspondence
with measures $s$ of the form

\begin{eqnarray}\label{e:nonkolmogorov}
&s:\mbox{$\mathcal{P}(\mathcal{H})$}\rightarrow[0,1]&\nonumber\\
&\noindent \mbox{such that}&\nonumber \\
\label{e:Qprobability1}
&s(\mathbf{0})= 0 \,\, (\mathbf{0}\,\, \mbox{is the null subspace}).&\nonumber\\
& s(P^{\bot})=1-s(P),&\\
&\noindent \mbox{and, for a denumerable}&\nonumber\\
&\mbox{and pairwise orthogonal family of projections}\,\,{P_{j}}\nonumber&\\
\label{e:Qprobability3}
&s(\sum_{j}P_{j})=\sum_{j}s(P_{j}).\nonumber&
\end{eqnarray}

\noindent Thus, given a state $\rho$, a measure $s_{\rho}$
satisfying Eqns. \ref{e:nonkolmogorov} is uniquely determined in
such a way that, for each outcome of each experiment represented by
a projection operator $P$, it coincides with the probability defined
in Feynman's quotation. In this way, probabilities appearing in QM
(which are governed by the density matrix and the Born's rule), can
be axiomatized using Eqs. \ref{e:nonkolmogorov}. How is all of this
related with the above Feynman's quotation? What is the technical
meaning of the radical difference mentioned by Feynman? While Eqs.
\ref{e:nonkolmogorov} may look unfamiliar, it is instructive to
consider a quantum probability distribution, such as $s$, {\it as a
collection of classical probability distributions}. Let us make some
important definitions in order to see how this works. Let
$E:=\{P_{i}\}_{i\in \mathbb{N}}$ be a collection of projections such
that $\bigvee_{i}P_{i}=\mathbf{1}_{\mathcal{H}}$ and $P_{i}\bot
P_{j}=0$ whenever $i\neq j$. We call $E$ an \emph{experiment}. The
intuitive idea of an experiment refers to the set of events defined
by a concrete experimental setup. Each one of these events is in a
bijective correspondence with a possible measurement outcome. Thus,
the $E$'s can be regarded as part of an outcome set $\Omega_{E}$. As
an example, measuring the spin of a particle in a definite direction
defines an experiment. To measure it in another direction, defines a
new experiment incompatible with the first one. Notice that any
experiment defines a maximal Boolean subalgebra of
$\mathcal{P}(\mathcal{H})$, which is isomorphic to
$\mbox{$\mathcal{P}$}(\Omega_{E})$. The state of the system defines
a classical probability distribution on this Boolean subalgebra,
satisfying Eqs. \ref{e:kolmogorovian}.

We call an orthonormal complete set of projectors of the form
$\{|\varphi_{i}\rangle\langle\varphi_{i}|\}_{i\in\mathbb{N}}$ in
$\mathcal{H}$ (where the $|\varphi_{i}\rangle$ are unit vectors) a
\emph{frame}. Notice that any frame is also an experiment. In a
sense, a frame represents a maximal experiment on the system, in the sense that it cannot be refined by any finer measurement (cfr. \cite{PeresBook}, Chapter 2). We
call $\mathcal{F}_{\mathcal{H}}$ to the set of all possible frames
in $\mathcal{H}$. Frames are irreducible experiments, in the sense that no outcome is degenerate.

Each experiment $E$ defines a maximal Boolean subalgebra
$\mathcal{B}_{E}\subset \mathcal{P}(\mathcal{H})$ \footnote{The
construction of $\mathcal{B}_{E}$ is trivial: it is indeed the
smallest Boolean subalgebra of $\mathcal{P}(\mathcal{H})$ containing
$E$.}. Again, if we restrict the state $\rho$ to $\mathcal{B}_{E}$,
we obtain a measure $\rho_{\mathcal{B}_{E}}$ on $\mathcal{B}_{E}$
satisfying Kolmogorov's probability theory (defined by Eqns.
\ref{e:kolmogorovian}).

Indeed, if we restrict to frames, for each orthonormal
basis $\{|\phi_{i}\rangle\}_{i\in\mathbb{N}}$ of $\mathcal{H}$
representing a particular irreducible experiment, the state $\rho$ assigns to
it a classical probability distribution represented by the vector
$(p_{|\phi_{1}\rangle},p_{|\phi_{2}\rangle},\ldots)$, where
$p_{|\phi_{i}\rangle}=\mbox{tr}(\rho|\phi_{i}\rangle\langle\phi_{i}|)$.
Indeed, the set
$\{|\phi_{i}\rangle\langle\phi_{i}|\}_{i\in\mathbb{N}}$ generates
a maximal Boolean algebra, and measure $s_{\rho}$ defines a
classical probability measure on it just as in
\ref{e:kolmogorovian}. Thus, the quantum probabilities
originated in a given state can be considered as a
(non-denumerable) family

\begin{eqnarray}
&\{(p_{|\phi_{1}\rangle},p_{|\phi_{2}\rangle},p_{|\phi_{3}\rangle},\ldots\ldots)\}&\nonumber\\
&\{(p_{|\phi'_{1}\rangle},p_{|\phi'_{2}\rangle},p_{|\phi'_{3}\rangle},\ldots\ldots)\}&\nonumber\\
&\{(p_{|\phi''_{1}\rangle},p_{|\phi''_{2}\rangle},p_{|\phi''_{3}\rangle},\ldots\ldots)\nonumber\}&\\
&\vdots&\nonumber\\
&\vdots&\nonumber\\
\end{eqnarray}

\noindent where $|\phi_{i}\rangle$, $|\phi'_{i}\rangle$, etc.,
ranges over all possible orthonormal basis of $\mathcal{H}$.

Thus, a quantum state can be seen as a collection of classical
probability distributions ranging over each possible experiment.
Since in QM different experiments can be incompatible (i.e., some of
them \emph{cannot be simultaneously performed}), a quantum state
does not determine a  single classical probability distribution: due
to Gleason's theorem, this fact is correctly axiomatized by Eqs.
\ref{e:nonkolmogorov}. We thus see how the meaning of the expression
``radically changed" in Feynman's quote can be expressed in a clear
technical (but also conceptual) sense. In classical probability
theory the rational agent is confronted with an event structure
represented by a single Boolean algebra (only one context). This is
the content of Cox' approach to probability theory \footnote{It is
also the content of other similar approaches as well, such as the
ones presented in Section II of
\cite{CavesFuchs-BayesianCoherence}}: the Boolean structure of
propositions representing classical events determine the possible
measures of degrees of belief. In other words, if the agent wants to
avoid inconsistencies, he must compute probabilities according to
rules compatible with the Boolean structure of classical logic.

In the quantum realm, due to the existence of complementary
contexts, a single Boolean algebra \emph{is no longer sufficient}
to cogently (and fully) describe physical phenomena, and thus, the
orthomodular structure of $\mathcal{P}(\mathcal{H})$ emerges. This
is the case for more general theories as well, such as algebraic
relativistic quantum field theory or quantum mechanics with
infinitely many degrees of freedom, and this involves the use of
more  general algebraic structures (more on this in the next
Section). Notice that these considerations do not imply that
classical logic should be abandoned; quite on the contrary, the
experimenter is always confronted with concrete experiments for
which a Boolean algebra is perfectly defined. But no a priori
principle grants that the complete description of all possible
phenomena will be exhausted within a single Boolean context. Here
we encounter  the radical difference in computing probabilities
that quantum mechanics forces on us: non-Boolean event structures
do appear in nature, and in this case, new rules for computing
probabilities must be invoked. In
\cite{Holik-Plastino-Saenz-AOP-2014} Cox' construction is
generalized by showing that when the experimenter is confronted
with events represented by a non-Boolean algebra such as
$\mathcal{P}(\mathcal{H})$, the plausibility measures must obey
Eqns. \ref{e:nonkolmogorov} in order to avoid inconsistencies.

\subsection{General case}\label{s:GeneralCase}

Measures in lattices more general than the sigma-algebra of the
classical case and $\mathcal{P}(\mathcal{H})$ can be constructed
\cite{RedeiSummersQuantumProbability,RedeiQuantumLogicInAlgebraicApproach}.
They can be axiomatized as conditions on a measure $s$ as follows:

\begin{eqnarray}\label{e:GeneralizedProbability}
&s:\mbox{$\mathcal{L}$}\rightarrow [0;1],&\nonumber\\
&(\mathcal{L}\,\,\mbox{standing for the lattice of all events)}&\nonumber\\
&\mbox{such that}&\nonumber\\
&s(\textbf{0})=0.&\nonumber\\
&s(E^{\bot})=1-s(E),&\\
&\mbox{and, for a denumerable and pairwise orthogonal family of events}\,E_{j}&\nonumber\\
&s(\sum_{j}E_{j})=\sum_{j}s(E_{j}).&\nonumber
\end{eqnarray}

\noindent See \cite{BeltramettiCassinelliBook} regarding the conditions for
the existence of such measures. Eqs. \ref{e:kolmogorovian} and
\ref{e:nonkolmogorov} are just particular cases of this general
approach. There do exist  concrete examples of measures on
lattices,  coming from Type II and Type III factors, which do not
reduce to \ref{e:kolmogorovian} and \ref{e:nonkolmogorov}
\cite{RedeiSummersQuantumProbability,Yngvason2005-TypeIIIFactors}.

Define an \emph{experiment} as a set of propositions
$\mathbf{A}:=\{a_{i}\}_{i\in\mathbb{N}}$, such that $a_{i}\perp
a_{j}$ for $i\neq j$ and $\bigvee_{i}a_{i}=\mathbf{1}$. Call
$\mathfrak{E}$ to  the set of all possible experiments. A frame in
$\mathcal{L}$ will be an orthogonal set
$\{a_{i}\}_{i\in\mathbb{N}}$ of atoms such that
$\bigvee_{i}a_{i}=\mathbf{1}$. Notice that frames are also
experiments here.

\section{Cox' Approach and Information
Measures}\label{s:CoxAndInformation}

Given an event structure (i.e, a set of propositions referring to events) represented by an atomic Boolean lattice $\mathcal{B}$, Cox defines a question as the set of assertions that answer it. If a proposition $x\in\mathcal{B}$ answers question $Q$ (notice that according to Cox' definition this means $x\in Q$), and if $y$ implies $x$ (or in lattice theoretical notation: $y\leq x$), then, $y$ should also answer $Q$ (and thus, $y\in Q$). Any set of propositions in $\mathcal{B}$ with this property will be called a \emph{down-set} (see \cite{Knuth-Neurocomputing-2005}). Thus, any question $Q$ in the set of questions $\mathcal{Q}(\mathcal{B})$ defined by $\mathcal{B}$ is a down-set. $\mathcal{Q}(\mathcal{B})$ forms a lattice with set theoretical inclusion as partial order, intersection as conjunction and set union as disjunction. Notwithstanding, $\mathcal{Q}(\mathcal{B})$ will fail to be Boolean, due to the failure of orthocomplementation.

Following \cite{Birkhoff-LatticeTheoryBook}, define  an
\emph{ideal} $I$ of a lattice $\mathcal{L}$ as a non-empty subset
satisfying the following conditions

\begin{itemize}
\item If $x\leq y$ and $y\in I$, then $x\in I$.
\item If $x,y\in I$, then $x\vee y\in I$.
\end{itemize}

\noindent Thus, any ideal is also a down-set. Given an element $a\in\mathcal{L}$, a set of the form
$I(a)=\{x\in\mbox{$\mathcal{L}$}\,\,|\,\,x\leq a\}$ is an ideal,
and it is called a \emph{principal ideal} of $\mathcal{L}$. An
important theorem due to Birkhoff \cite{Birkhoff-LatticeTheoryBook} asserts that the set $\hat{\mathcal{L}}$ of all
ideals forms a lattice and the set $\hat{\mathcal{L}}_{p}$ of all
principal ideals forms a sublattice, which is isomorphic to
$\mathcal{L}$ \cite{Birkhoff-LatticeTheoryBook} (and we denote
this fact by $\hat{\mathcal{L}}_{p}\sim\mathcal{L}$).

For an arbitrary
atomic Boolean algebra $\mathcal{B}$, any $a\in\mathcal{B}$ can be written in the form
$a=\bigvee_{i}a_{i}$, for some atoms $a_{i}$\footnote{Notice
that maximal Boolean subalgebras of $\mathcal{P}(\mathcal{H})$
satisfy these conditions and that the disjunction may be infinite
but denumerable.}. We can also form the
lattices of ideals $\hat{\mathcal{B}}$ and
$\hat{\mathcal{B}}_{p}$, with
$\hat{\mathcal{B}}_{p}\subseteq\hat{\mathcal{B}}$ and
$\hat{\mathcal{B}}_{p}\sim\mathcal{B}$ (as lattices).

We can also form the lattice of questions $\mathcal{Q}(\mathcal{B})$ (which will
not be necessarily suitably orthocomplemented). Notice that while each ideal in
$\mathcal{B}$ belongs to $\mathcal{Q}(\mathcal{B})$, not every
element in $\mathcal{Q}(\mathcal{B})$ is an ideal (because a
system of assertions does not necessarily satisfy the join
condition of the definition of ideal \cite{KnuthWhatIsAQuestion}).
Thus, in order to stress the difference, let us call
$\hat{\mathcal{Q}}(\mathcal{B})$ to the set of ideal-questions
(i.e., questions such that are represented by ideals of
$\mathcal{B}$). It should be clear that
$\hat{\mathcal{Q}}(\mathcal{B})\subseteq\mathcal{Q}(\mathcal{B})$.
For any question $Q\in\mathcal{Q}(\mathcal{B})$, if $a\in
Q$, then, the ideal $I(a)$ of $a$ in $\mathcal{B}$ satisfies
$I(a)\subseteq Q$ (because $Q$ must contain all the $x$ such that
$x\leq a$). From this, it follows that $Q=\bigcup_{a\in Q}I(a)$.

One more step is needed in order to guarantee that our questions
be \emph{real}. A real question must satisfy the condition of
being answerable by a true statement \cite{Knuth-Neurocomputing-2005}. This is elegantly done by
requiring that all atoms must belong to a question in order to be
considered real. Thus, let $\mathcal{R}(\mathcal{B})$ be the set
if real questions. It can be shown that in the general case,
$\mathcal{R}_{\mathcal{B}}$ will not be Boolean because of the
failure of orthocomplementation. We will not use this lattice
here, but only consider $\mathcal{Q}(\mathcal{B})$ and
$\hat{\mathcal{Q}}(\mathcal{B})$.

There exists a quantity analogous to probability, called relevance
\cite{Knuth-Neurocomputing-2005}, which quantifies the degree to
which one question answers another (the technical details of the
construction of the relevance function are similar to those
presented in Section $2.2$). Relevance is not
only a natural generalization of information theory, but also
forms its foundation \cite{Knuth-Neurocomputing-2005}. Let us
repeat that the vocable relevance refers to the computation of to
what an extent a question answers another one. From the
mathematical point of view, this task is completely analogous to
that of assigning plausibility to $\mathcal{B}$, but applied now
to $\mathcal{Q}(\mathcal{B})$. As explained in
\cite{Knuth-Neurocomputing-2005}, in order to assign relevances,
i) the algebraic properties of the question lattice
$\mathcal{Q}(\mathcal{B})$ and ii) the probability assigned to
$\mathcal{B}$ using Cox' method must be taken into account. The
objective is thus to assign relevances to the ideal-questions (the
rest can be computed using the inquiry calculus derived using Cox'
method, see Knuth \cite{Knuth-Neurocomputing-2005}). With the
question algebra well-defined, Knuth  extends the ordering
relation to a quantity that describes the degree to which one
question answers another. This is done by defining a bi-valuation
on the lattice that takes two questions and returns a real number
$d \in [0, c]$, where $c$ is the maximal relevance. Precisely,
Knuth calls this bi-valuation the relevance
\cite{Knuth-Neurocomputing-2005}. This procedure can be applied to
$\mathcal{Q}(\mathcal{B})$ and $\hat{\mathcal{Q}}(\mathcal{B})$,
and thus we have a function $d(\cdot|\cdot)$ with properties
analogous to that of a plausibility function, but now defined on
the lattices of questions.

Following \cite{Knuth-Neurocomputing-2005}, we assume that the
extent to which the top question  $\mathbf{\hat{1}}$ answers a
join-irreducible question $I(a_{i})$ depends only on the probability
of the assertion $a_{i}$ from which the question $I(a_{i})$ was
generated. More abstractly,
$d(I(a_{i})|\mathbf{\hat{1}})=H(p(a_{i})|\mathbf{1})$,  $H$ being a
function to be determined in such a way that it satisfies
compatibility with the algebraic properties of the lattice and the
probabilities assigned in $\mathcal{B}$ (by using Cox' method). Now,
let us review the properties of  $d(\cdot|\cdot)$ according to
Knuth' inquiry calculus. First, we will have subadditivity

\begin{equation}
d(a\vee b|c)\leq d(a|c)+d(b|c)
\end{equation}
\noindent which is a straightforward consequence of the
sigma-additivity condition

\begin{equation}
d(\bigvee_{i}x_{i}|c)=\sum_{i}d(x_{i}|c)
\end{equation}

\noindent for pairwise disjoint questions
$\{x_{i}\}_{i\in\mathbb{N}}$. Commutativity of ``$\vee$" implies
that

\begin{equation}
d(x_{1}\vee x_{2}\vee\ldots\vee x_{n}|c)=d(x_{\pi(1)}\vee
x_{\pi(2)},\ldots\vee x_{\pi(n)}|c)
\end{equation}

\noindent for any permutation $\pi$. Now suppose that to a certain collection of questions $\{x_{1},x_{2},\ldots,x_{n}\}$ we add a new question $y=I(x)$ and that we know in advance that the assertion $x$ is false. Then, $y$ collapses to $\mathbf{\hat{0}}\in\mathcal{Q}(\mathcal{B})$. Thus, we should have the \emph{expansibility} condition

\begin{equation}
d(x_{1}\vee x_{2}\vee\ldots\vee x_{n}\vee y|c)=d(x_{1}\vee
x_{2}\vee\ldots\vee x_{n}|c)
\end{equation}

\noindent Suppose now that a question $X$ in $\mathcal{Q}(\mathcal{B})$ can be written as $X=\bigvee_{i}I(a_{i})$, where the $\{I(a_{i})\}$ are ideal questions with $I(a_{i})\wedge I(a_{j})=\mathbf{\hat{0}}$. Then, we will have

\begin{equation}
d(\bigvee_{i}I(a_{i})|\mathbf{\hat{1}})=\sum_{i}d(\bigvee_{i}I(a_{i})|\mathbf{\hat{1}})=\sum_{i}H(p(a_{i})|\mathbf{1}).
\end{equation}

\noindent Let us cast the above equation as

\begin{equation}
d(\bigvee_{i}I(a_{i})|\mathbf{\hat{1}})=K(p(a_{i})),
\end{equation}

\noindent where we have introduced the function $K(p(a_{i}))$
which depends on the $p(a_{i})$ only. If the $\{I(a_{i})\}$ form a
finite set (of $n$ elements), we can write
$K(p(a_{i}))=K_{n}(p(a_{1}),\ldots,p(a_{n}))$. It turns out that
$K_{n}(p(a_{1}),\ldots,p(a_{n}))$ satisfies subadditivity,
additivity, symmetry and expansibility. A well known result
\cite{Aczel-WhyShannon-1974,Knuth-Neurocomputing-2005} implies that

\begin{equation}\label{e:GeneralFormEntropy}
K_{n}(p(a_{1}),\ldots,p(a_{n}))=A H_{n}(p(a_{1}),\ldots,p(a_{n}))+ B
H_{n}^{0}(p(a_{1}),\ldots,p(a_{n})),
\end{equation}

\noindent where $A$ and $B$ are arbitrary constants,
$H_{n}(p_{1},\ldots,p_{n})=-\sum_{i=1}^{n}p_{i}\ln p_{i}$ and
$H_{n}^{0}=\ln(n)$ are the Shannon and Hartley entropies
respectively. For information theoretical purposes related to the continuity of the measure of information \cite{Aczel-WhyShannon-1974,Knuth-Neurocomputing-2005}, it is very natural to set $A=1$ and $B=0$, and thus
$K_{n}(p(a_{1}),\ldots,p(a_{n}))=-\sum_{i=1}^{n}p(a_{i})\ln
p(a_{i})$. When the terms $\{I(a_{i})\}$ in the decomposition are
an infinite denumerable set, by continuity, we will have that
$K(p(a_{i}))=-\sum_{i=1}^{\infty}p(a_{i})\ln p(a_{i})$. The
discussion in this Section allows us to discard the restriction to
finite Boolean algebras and turn to more general ones.

\section{Von Neumann's Entropy As A Natural Measure For
$\mathcal{P}(\mathcal{H})$}\label{s:vNeumannAsANaturalMeasure}

As was done in the Cox approach to the Boolean case in order to justify the use of Shannon measure, we look now for a natural information measure for $\mathcal{P}(\mathcal{H})$, i.e., a function depending on the non-commutative measure defined by Eqns. \ref{e:nonkolmogorov}. In other words, by appealing to Gleason's theorem, we look for a function $S(\rho)$ (depending \emph{only} on the state $\rho$), and at the same time compatible with the algebraic structure of $\mathcal{P}(\mathcal{H})$. Notice that it is not a priori obvious whether a variant of Cox method can be applied to the non-Boolean structure of $\mathcal{P}(\mathcal{H})$ and used to justify the choice of the VNE.  In this Section we will see that, according to Cox approach, \emph{the VNE appears as the most rational choice}.

Let us call $\mathfrak{B}_{\mathcal{P}(\mathcal{H})}$ to the set of
all maximal Boolean lattices of $\mathcal{P}(\mathcal{H})$. For each
$\mathcal{B}\in\mathfrak{B}_{\mathcal{P}(\mathcal{H})}$, we can
consider its dual lattice of ideals $\hat{\mathcal{B}}$.

Notice that when $\mathcal{H}$ is finite dimensional, its maximal
Boolean subalgebras will be finite. As an example, consider
$\mathcal{P}$($\mathbb{C}^{2}$), i.e., the set of all possible
linear subspaces of a two dimensional complex Hilbert space. Then,
each maximal Boolean subalgebra will be of the form
$\{\mathbf{0},\mathbf{P},\neg\mathbf{P}^{\bot},\mathbf{1}_{\mathbb{C}^{2}}\}$,
with $\mathbf{P}=|\varphi\rangle\langle\varphi|$ for some unit norm
vector $|\varphi\rangle$ and
$\mathbf{P}^{\bot}=|\varphi^{\bot}\rangle\langle\varphi^{\bot}|$
(with $\langle\varphi|\varphi^{\bot}\rangle=0$). In a similar way,
for $\mathcal{P}$($\mathbb{C}^{3}$), a maximal Boolean subalgebra
will be isomorphic to
$\mathcal{P}(\{$a,b,c$\})=\{\emptyset,\{$a$\},\{$b$\},\{$c$\},\{$a,b$\},\{$a,c$\},\\\{$b,c$\},\{$a,b,c$\}\}$.
More specifically, for this last example, given three orthogonal
rays in $\mathbb{C}^{3}$ represented by unitary vectors
$|\varphi_1\rangle$, $|\varphi_2\rangle$ and $|\varphi_3\rangle$,
the set
$\{\mathbf{0},\mathbf{P}_{1},\mathbf{P}_{2},\mathbf{P}_{3},\mathbf{P}_{12},\mathbf{P}_{13},\mathbf{P}_{23},\mathbf{1}_{\mathbb{C}^{3}}\}$,
where $P_{i}=|\varphi_i\rangle\langle\varphi_i|$ ($i=1,2,3$) and
$P_{ij}=|\varphi_i\rangle\langle\varphi_i|+|\varphi_j\rangle\langle\varphi_j|$
($i,j=1,2,3$, $i\neq j$), forms a maximal Boolean subalgebra (and
all maximal Boolean subalgebras are of this form). Notice that in
these examples, the sets of atoms
$\{|\varphi_1\rangle\langle\varphi_1|;|\varphi_2\rangle\langle\varphi_2|;|\varphi_3\rangle\langle\varphi_3|\}$
(with orthonormal $|\varphi_i\rangle\langle\varphi_i|$ for all $i$)
and
$\{|\varphi\rangle\langle\varphi|;|\varphi^{\bot}\rangle\langle\varphi^{\bot}|\}$
i) form frames, and ii) generate the above mentioned Boolean
subalgebras of $\mathcal{P}(\mathcal{H})$.

Now, it is important to notice that if we restrict a state $\rho$ to
$\mathcal{B}$, we will have a classical probability measure such as
the one defined by Eqns. \ref{e:kolmogorovian}, and a concomitant
inquiry set $\mathcal{Q}(\mathcal{B})$ can be defined as in
\cite{CoxLibro,Knuth-DerivingLaws} (see Section
$3$ of this work). In what follows, our
strategy will be to construct a suitable information measure, just
as we did in Section $3$, for each maximal
Boolean subalgebra of $\mathcal{P}(\mathcal{H})$. For each frame
$F=\{|\varphi_{i}\rangle\langle\varphi_{i}|\}_{i\in\mathbb{N}}\subset
\mathcal{B}$ representing a complete experiment, state $\rho$
assigns probabilities
$p_i=\mbox{tr}(\rho|\varphi_{i}\rangle\langle\varphi_{i}|)$ to each
possible outcome of $F$. By following Cox' spirit
\cite{CoxLibro,Knuth-DerivingLaws,Knuth-MeasuringQuestions} and the
procedure sketched in Section $3$, we can
guarantee (by choosing suitable coefficients $A$ and $B$ in Eqn.
\ref{e:GeneralFormEntropy}) that for each maximal Boolean subalgebra
$\mathcal{B}$ there exists a canonical information measure
$H_{F}(\rho)$ such that for each frame $F\subseteq\mathcal{B}$:

\begin{eqnarray}\label{e:CanonicalMeasureInGeneralFrame}
H_{F}(\rho)=-\sum_{i}p_i\ln
p_i=\nonumber\\-\sum_{i}\mbox{tr}(\rho|\varphi_{i}\rangle\langle\varphi_{i}|)\ln(\mbox{tr}(\rho|\varphi_{i}\rangle\langle\varphi_{i}|)).
\end{eqnarray}

The above construction can be carried out for any
$\mathcal{B}\in\mathfrak{B}_{\mathcal{P}(\mathcal{H})}$. Thus, for any $\rho$, each
$\mathcal{B}\in\mathfrak{B}_{\mathcal{P}(\mathcal{H})}$ and each frame $F\subseteq\mathcal{B}$, we have a measure $H_{F}(\rho)$.
It is important to note that this family of measures, although only
defined on the maximal boolean sublattices, do cover the whole
$\mathcal{P}(\mathcal{H})$ lattice. This is so because, as shown
in \cite{navara1991pasting}, every orthomodular lattice is the
union of its maximal boolean sublattices.

Our point is that we need a measure such that it depends only on
$\rho$ and not on the particular choice of complete experiment
(represented by a particular frame). Among the family  of measures
$H_{F}(\rho)$, it is natural (according to Cox approach) to take the
one which attains the minimum value: the one with the least
Shannon's information (i.e., we are looking for the frame in which
the information is maximal). This means that it is natural to define

\begin{equation}
H(\rho):=\inf_{F\in\mathcal{F}_{\mathcal{H}}}H_{F}(\rho).
\end{equation}

\noindent Given that $\rho$ is self adjoint, let us consider its set of eigenprojectors
$F_{\rho}=\{|\rho_{i}\rangle\langle\rho_{i}|\}_{i\in\mathbb{N}}$,
with $\rho_{i}\in\mathbb{R}$ satisfying
$\rho|\rho_{i}\rangle=\rho_{i}|\rho_{i}\rangle$ and
$\rho=\sum\rho_{i}|\rho_{i}\rangle\langle\rho_{i}|$.
It should be clear that if $\rho$ is non-degenerate, $F_{\rho}$ is
a frame. If $\rho$ is degenerate, it is equally easy to find a
frame out of its eigenprojections. Accordingly,  without loss of
generality we can suppose that $\rho$ defines a frame. Now
consider the maximal Boolean algebra $\mathcal{B}_{F_{\rho}}$
generated by $F_{\rho}$. Using Eq.
\ref{e:CanonicalMeasureInGeneralFrame}, it follows that the
canonical measure $H$, when restricted to $F_{\rho}$ satisfies

\begin{eqnarray}
&H_{F_{\rho}}(\rho)=-\sum_{i}\mbox{tr}(\rho|\rho_{i}\rangle\langle\rho_{i}|)\ln(\mbox{tr}(\rho|\rho_{i}\rangle\langle\rho_{i}|))=\nonumber\\
&-\sum_{i}\rho_{i}\ln\rho_{i}=-\mbox{tr}(\rho\ln(\rho))
\end{eqnarray}

\noindent which is nothing but the VNE. But the VNE has the well known
property of attaining its minimum value at $F_{\rho}$ (cf. Reference \cite{Bengtsson2006}):

\begin{equation}
-\mbox{tr}(\rho\ln(\rho))\leq H_{F}(\rho), \,\,\, \forall
\,\,\,F\in\mathcal{F}_{\mathcal{H}}
\end{equation}

\noindent Thus, we have shown that $H(\rho)=-\mbox{tr}(\rho\ln(\rho))$. In other words, von Neumann's entropy is the only function which emerges canonically
as the minimum of all measures compatible with the algebraic
structure of $\mathcal{P}(\mathcal{H})$. Notice that we are
\emph{deriving} VNE out of the algebraic symmetries of the
lattice. The above considerations show the VNE as a natural
measure of information of $\mathcal{P}(\mathcal{H})$, as a
consequence of Shannon's entropy being the natural information
measure of a Boolean algebra following Cox' method. Notice that our derivation covers both the finite and infinite dimensional cases.

\section{Generalized Probabilistic
Models}\label{s:GeneraliedProbabilisticModels}

After deriving the VNE using the Cox method, we now advance a step further and investigate whether this procedure can be extended to more general contextual theories. Concretely, we now briefly discuss what happens if $\mathcal{L}$ is an arbitrary atomic orthomodular lattice and $\mu$ is a measure obeying Eqs. \ref{e:GeneralizedProbability}. We show that the procedure of the previous Section can be extended to this case. Let
$\mathfrak{B}_{\mathcal{L}}$ be the set of all possible maximal
Boolean subalgebras of $\mathcal{L}$. For each
$\mathcal{B}\in\mathfrak{B}_{\mathcal{L}}$, the Cox' construction
applies as in Section $3$, and we have a
Shannon's function $H_{\mathbf{F}}(\mu)$ defined for each frame
$\mathbf{F}=\{a_{i}\}_{i\in\mathbb{N}}\in\mathfrak{E}$ (see
Section $2.4$):

\begin{equation}
H_{\mathbf{F}}(\mu)=-\sum_{a_{i}\in A}\mu(a_{i})\ln(\mu(a_{i})),
\end{equation}

\noindent As in the previous Section, we define:

\begin{equation}\label{e:MeasurementEntropy}
H(\mu):=\inf_{\mathbf{F}\in\mathfrak{E}}H_{\mathbf{F}}(\mu).
\end{equation}

\noindent Notice that when restricted to frames,
$H_{\mathbf{A}}(\mu)$ coincides with the Shannon's measures
derived using Cox' method. Thus, by construction, $H(\mu)$ does
the job of representing the canonical measure of information, as
Shannon's and VNE did in the classical and quantum cases,
respectively.

The results of this Section show that it is indeed possible to
generalize Cox' method to probabilistic theories more general than a
Boolean algebra. Notice that, when $\mathcal{L}$ is a Boolean
algebra, we recover Cox' construction, and when
$\mathcal{L}=\mathcal{P}(\mathcal{H})$, we recover our construction
for the VNE. Indeed, by looking at Eq. \ref{e:MeasurementEntropy}, the reader will soon recognize that our derivation coincides with the \emph{measurement entropy} (ME) introduced in \cite{Short-Wehner-2010,Barnum2010,HeinEntropyInQL1979}. The main difference of our approach with the one of these references is that: i) we derive the same measures by using Cox approach, and thus, we provide a novel intuitive interpretation for them; and ii) by means of our derivation, we discard other possible functional forms, such as the ones appearing in Tsallis or R\'{e}nyi entropies, justifying in this way the usage of the logarithmic form of the VNE and the ME.

\section{Conclusions}\label{s:Conclusions}

If a rational agent deals with a Boolean algebra of
assertions, representing physical events, a plausibility calculus can be derived in such a way that the plausibility function yields a theory which is formally
equivalent to that of Kolmogorov for classical probabilities
\cite{Cox-ProbabilityFrequency-1946,CoxLibro,Knuth-Skilling-Axioms-2012,Knuth-Neurocomputing-2005}.

A similar result holds if the rational agent deals with an atomic orthomodular lattice \cite{Holik-Plastino-Saenz-AOP-2014}, as is the
case with the contextual character of the lattice of projections
representing events of a quantum system. For the later case,
non-Kolmogorovian probabilities (Eqs. \ref{e:nonkolmogorov})
arise as the only ones compatible with the non-commutative (non-Boolean) character of quantum complementarity.

In Cox' approach,  Shannon's information measure relies on the
axiomatic structure of Kolmogorovian probability theory. We have
shown  in Section $4$  that,
according to our extension of Cox' method, the VNE emerges as its
non-commutative version. The VNE thus arises as a natural measure
of information derived from the non-Boolean character of the
underlying lattice $\mathcal{P}(\mathcal{H})$. The different
entropies discussed in this work are summarized in Table $1$.

The fact that this kind of construction can be extended to more
general probabilistic models (as we have shown in Section
$5$, where we have deduced the
ME as a natural measure of information), implies that CIT and QIT can be considered as  particular cases of a more general
\emph{non-commutative information theory}.

These results allow for an interpretation of the VNE and measurement
entropy as the natural measures of information for an experimenter
who deals with a non-Boolean (contextual) event structure. This is
the case for standard quantum mechanics, in which quantum
complementarity expresses itself in the existence of non-compatible
measurement set-ups and, consequently, in the different contexts of
$\mathcal{P}(\mathcal{H})$ (maximal Boolean subalgebras) and
non-commutative observables.


\begin{table}[h]\label{t:Table1}
\centering
\begin{tabular}{l|l|l|l|}
\cline{2-4} & \,\,\,\,\,\,\,\,\,\textsc{Classical} &
\,\textsc{Quantum} & \,\,\,\,\,\textsc{General}
\\ \hline
\multicolumn{1}{|l|}{\textsc{Lattice}} &
\,\,\,\,\,\,\,\,\,\,\,\,\,\,\,$\mbox{$\mathcal{P}$}(\Gamma)$ &
\,\,\,\,\,\,\,\,\,$\mathcal{P}(\mathcal{H})$ &
\,\,\,\,\,\,\,\,\,\,\,\,\,$\mathcal{L}$
\\ \hline
\multicolumn{1}{|l|}{\textsc{Entropy}} & $-\sum_{i}p(i)\ln(p(i))$ &
$-\mbox{tr}\rho\ln(\rho)$ &
$\inf_{\mathbf{F}\in\mathfrak{E}}H_{\mathbf{F}}(\mu)$
\\ \hline
\end{tabular}
\vspace{0.2cm}
\caption{Table comparing the differences between the classical,
quantal, and general cases.}
\end{table}

\vskip1truecm

\noindent {\bf Acknowledgements} \noindent We acknowledge CONICET and UNLP. We thank to the anonymous Referees for useful comments and suggestions.

\appendix

\noindent
\emph{Lattices}

\begin{itemize}
\item A lattice $\mathcal{L}$ is a partially ordered set (i.e., a set
endowed with a partial order relationship ``$\leq$") such that for
very $a,b\in\mathcal{L}$ there exists a unique supremum, the least
upper bound ``$a\vee b$" called their \emph{join}, and an infimum,
the greatest lower bound ``$a\wedge b$" called their \emph{meet}.

\item A bounded lattice has a greatest and least element, denoted $\mathbf{1}$
and $\mathbf{0}$ (also called \emph{top} and \emph{bottom},
respectively).

\item For any lattice, an orthocomplementation is a
unary operation ``$\neg(\ldots)$" satisfying:

\begin{subequations}\label{e:ComplementationAxioms}
\begin{equation}\label{e:Complement1}
\neg(\neg(a))=a
\end{equation}
\begin{equation}\label{e:Complement2}
a\leq b \longrightarrow \neg b\leq \neg a
\end{equation}
$a\vee \neg a$ and $a\wedge \neg a$ exist and
\begin{equation}\label{e:Complement3}
a\vee \neg a=\mathbf{1}
\end{equation}
\begin{equation}\label{e:Complement4}
a\wedge \neg a=\mathbf{0}
\end{equation}
hold.
\end{subequations}

\item If $\mathcal{L}$ has a null element $ 0$, then an element $x$ of $\mathcal{L}$
is an \emph{atom} if $0 < x$ and there exists no element $y$ of
$\mathcal{L}$ such that $0 < y < x$. $\mathcal{L}$ is \emph{Atomic}, if for every nonzero element $x$ of
$\mathcal{L}$, there exists an atom $a$ of $\mathcal{L}$ such that $
a \leq x$.

\item A \emph{modular} lattice is one that satisfies the modular law
$x \leq b$ implies $x \vee (a \wedge b) = (x \vee a) \wedge b$,
where $\le$ is the partial order, and $\vee$ and $\wedge$ (join and
meet, respectively) are the operations of the lattice. An \textit{orthomodular lattice} is an orthocomplemented lattice satisfying the orthomodular law: $a \leq b$ and $\neg a\leq
c$ implies $a \vee (b \wedge c) = (a \vee b) \wedge (a\vee c)$.

\item \emph{Distributive} lattices are lattices for which the operations of join
and meet are distributive over each other. Distributive
orthocomplemented lattices are called \emph{Boolean}. The collection
of subsets of a given set, with set intersection as meet, set union
as join and set complement as orthocomplementation, form a complete
bounded lattice which is also Boolean.

\item Any quantum system represented by a separable Hilbert space
$\mathcal{H}$ has associated a lattice formed by all its closed
subspaces $\mathcal{P}(\mathcal{H})$, where
$\mathbf{0}$ is the null subspace, $\mathbf{1}$ is the total space
$\mathcal{H}$, $\vee$ is the closure of the direct sum, $\wedge$ is subspace intersection, and $\neg(S)$ is the
orthogonal complement of a subspace $S^{\bot}$
\cite{RedeiQuantumLogicInAlgebraicApproach}. This lattice was called
``Quantum Logic" by Birkhoff and von Neumann and it is a modular one
if the Hilbert space is finite dimensional, and orthomodular for the
infinite dimensional case. The set of projection operators on
$\mathcal{H}$ forms a lattice which is isomorphic to
$\mathcal{P}(\mathcal{H})$ (and thus, they can be identified).

\end{itemize}


\noindent
\textbf{References}
\begin{thebibliography}{00}

\bibitem{Tribus71}
M.~Tribus and E.~C. Mcirvine. Energy and Information, Sci. Am., 225(3):179--188, 1971.

\bibitem{Bengtsson2006} I. Bengtsson and K. \.{Z}yczkowski. \textit{Geometry of Quantum States: An Introduction to Quantum Entanglement} (Cambridge University Press, Cambridge, 2006).

\bibitem{NielsenBook}
Michael~A Nielsen and Isaac~L Chuang. Quantum computation and quantum information, Cambridge university press, 2010.

\bibitem{Schumacher-QuantumCoding-1995}
Benjamin Schumacher. Quantum coding. Phys. Rev. A, 51:2738--2747, Apr 1995.

\bibitem{Jozsa-Schumacher-NewProof-1994}
Richard Jozsa and Benjamin Schumacher. A new proof of the quantum noiseless coding theorem. Journal of Modern Optics, 41(12):2343--2349, 1994.

\bibitem{TsallisEntropy1988}
Constantino Tsallis.
 Possible generalization of boltzmann gibbs statistics.
  Journal of Statistical Physics, 52(1-2):479--487, 1988.

\bibitem{RasteginQIC-2014}
Alexey Rastegin.
 Tests for quantum contextuality in terms of q-entropies.
  Quantum Information And Computation, 14:0996--1013, September 2014.

\bibitem{renyi1961}
Alfr\'{e}d R\'{e}nyi.
  On Measures of Entropy and Information.
 University of California Press, Berkeley, Calif., 1961.

\bibitem{Rossignoli-2010}
R.~Rossignoli, N.~Canosa, and L.~Ciliberti.
 Generalized entropic measures of quantum correlations.
  Phys. Rev. A, 82:052342, Nov 2010.

\bibitem{Zozor-Bosyk-Portesi-2014}
S~Zozor, G~M Bosyk, and M~Portesi.
 General entropy-like uncertainty relations in finite dimensions.
  Journal of Physics A: Mathematical and Theoretical,
  47(49):495302, 2014.

\bibitem{MullerRenyi}
M.~M\"{u}ller-Lennert, F.~Dupuis, O.~Szehr, S.~Fehr, and M.~Tomamichel.
 On quantum r\'{e}nyi entropies: A new generalization and some
  properties.
  Journal of Mathematical Physics, 54(12), 2013.

\bibitem{Zeilinger-AgainstShannon-2001}
\ifmmode \check{C}\else~\v{C}\fi{}aslav Brukner and Anton Zeilinger.
 Conceptual inadequacy of the shannon information in quantum
  measurements.
  Phys. Rev. A, 63:022113, Jan 2001.

\bibitem{Timpson-AgainstZeilinger}
C.G. Timpson.
 On a supposed conceptual inadequacy of the shannon information in
  quantum mechanics.
  Studies in History and Philosophy of Science Part B: Studies in
  History and Philosophy of Modern Physics, 34(3):441 -- 468, 2003.
 Quantum Information and Computation.

\bibitem{Aczel-WhyShannon-1974}
J.~Acz{\'e}l, B.~Forte, and C.~T. Ng.
 Why the {S}hannon and {H}artley entropies are `natural'.
  Advances in Appl. Probability, 6:131--146, 1974.

\bibitem{AczelBook-1966}
J.~Acz{\'e}l.
  Lectures on functional equations and their applications.
 Mathematics in Science and Engineering, Vol. 19. Academic Press, New
  York-London, 1966.
 Translated by Scripta Technica, Inc. Supplemented by the author.  Edited by Hansjorg Oser.

\bibitem{Ochs-UniqueCharacterization-1975}
W.~Ochs.
 A unique characterization of the generalized
  {B}oltzmann-{G}ibbs-{S}hannon entropy.
  Phys. Lett. A, 54(3):189--190, 1975.

\bibitem{Ochs-UniqueCharacterization-1976}
W.~Ochs.
 A unique characterization of the generalized
  {B}oltzmann-{G}ibbs-{S}hannon entropy.
  Rep. Mathematical Phys., 9(3):331--354, 1976.

\bibitem{Ochs-VN-1975}
W.~Ochs.
 A new axiomatic characterization of the von {N}eumann entropy.
  Rep. Mathematical Phys., 8(1):109--120, 1975.

\bibitem{Shannon1948}
Claude~E Shannon.
 A mathematical theory of communication, part i.
  Bell Syst. Tech. J., 27:379--423, 1948.

\bibitem{KolmogorovProbability}
A.N. Kolmogorov.
  Foundations of Probability Theory.
 Julius Springer: Berlin, Germany, 1933.

\bibitem{Cox-ProbabilityFrequency-1946}
R.~T. Cox.
 Probability, frequency and reasonable expectation.
  American Journal of Physics, 14(1), 1946.

\bibitem{CoxLibro}
R.T. Cox.
  The Algebra of Probable Inference.
 The Johns Hopkins Press: Baltimore, MD, USA, 1961.

\bibitem{Knuth-DerivingLaws}
Kevin~H. Knuth.
 Deriving laws from ordering relations.
 In  Bayesian inference and maximum entropy methods in science and
  engineering, volume 707 of  AIP Conf. Proc., pages 204--235. Amer.
  Inst. Phys., Melville, NY, 2004.

\bibitem{Goyal-Knuth-Symmetry}
Philip Goyal and Kevin~H. Knuth.
 Quantum theory and probability theory: their relationship and origin
  in symmetry.
  Symmetry, 3(2):171--206, 2011.

\bibitem{Knuth-Neurocomputing-2005}
Kevin~H. Knuth.
 Lattice duality: The origin of probability and entropy.
  Neurocomputing, 67(0):245 -- 274, 2005.
 Geometrical Methods in Neural Networks and Learning.

\bibitem{Knuth-MeasuringQuestions}
Kevin~H. Knuth.
 Measuring questions: relevance and its relation to entropy.
 In  Bayesian inference and maximum entropy methods in science and engineering, volume 735 of  AIP Conf. Proc., pages 517--524. Amer. Inst. Phys., Melville, NY, 2004.

\bibitem{knuth-aistats2005}
Kevin Knuth.  Toward question-asking machines: The logic of questions and the inquiry calculus.
 In Robert~G. Cowell and Zoubin Ghahramani, editors,  aistats05, pages 174--180. Society for Artificial Intelligence and Statistics, 2005.
 (Available electronically at http://www.gatsby.ucl.ac.uk/aistats/).

\bibitem{KnuthWhatIsAQuestion}
Kevin~H. Knuth.
 What is a question? In  Bayesian inference and maximum entropy methods in science and engineering, volume 659 of  AIP Conf. Proc., pages 227--242. Amer. Inst. Phys., Melville, NY, 2003.

\bibitem{Knuth-Skilling-Axioms-2012}
Kevin~H. Knuth and John Skilling. Foundations of inference. Axioms, 1(1):38--73, 2012.

\bibitem{Holik-Plastino-Saenz-AOP-2014}
Federico Holik, Manuel S\'{a}enz, and Angel Plastino. A discussion on the origin of quantum probabilities. Annals of Physics, 340(1):293 -- 310, 2014.

\bibitem{Short-Wehner-2010}
Anthony~J Short and Stephanie Wehner. Entropy in general physical theories. New Journal of Physics, 12(3):033023, 2010.

\bibitem{Barnum2010}
Howard Barnum, Jonathan Barrett, Lisa~Orloff Clark, Matthew Leifer,
Robert Spekkens, Nicholas Stepanik, Alex Wilce, and Robin Wilke. Entropy and information causality in general probabilistic theories.  New Journal of Physics, 12:033024, 2010.

\bibitem{HeinEntropyInQL1979}
Carl A. Hein.
 Entropy in operational statistics and quantum logic.
  Foundations of Physics, 9(9-10):751--786, 1979.

\bibitem{vN-OriginalPaperOnVNAlgebras-1930}
J.~v.~Neumann.
 Zur algebra der funktionaloperationen und theorie der normalen operatoren.
  Mathematische Annalen, 102(1):370--427, 1930.

\bibitem{RedeiSummersQuantumProbability}
Mikl{\'o}s R{\'e}dei and Stephen~Jeffrey Summers.
 Quantum probability theory.
  Studies in History and Philosophy of Science Part B: Studies in History and Philosophy of Modern Physics, 38(2):390 -- 417, 2007.

\bibitem{RedeiQuantumLogicInAlgebraicApproach}
Mikl{\'o}s R{\'e}dei.
  Quantum logic in algebraic approach, volume~91 of
  Fundamental Theories of Physics.
 Kluwer Academic Publishers Group, Dordrecht, 1998.

\bibitem{Murray-VN-OnRingsOfOperatorsI}
F.~J. Murray and J.~von Neumann.
 On rings of operators.
  Ann. Of Math., 37(1):116--229, 1936.

\bibitem{Murray-VN-OnRingsOfOperatorsII}
F.~J. Murray and J.~von Neumann.
 On rings of operators ii.
  Trans. Amer. Math. Soc., 41(1):208--248, 1937.

\bibitem{VN-OnRingsOfOperatorsIII}
J.~von Neumann.
 On rings of operators iii.
  Ann. Of Math., 41(1):94--161, 1940.

\bibitem{Murray-VN-OnRingsOfOperatorsIV}
F.~J. Murray and J.~von Neumann.
 On rings of operators iv.
  Ann. Of Math., 44(4):716--808, 1943.

\bibitem{Yngvason2005-TypeIIIFactors}
Jakob Yngvason.
 The role of type III factors in quantum field theory.
  Reports on Mathematical Physics, 55(1):135--147, 2005.

\bibitem{AcacioNegative}
J~Acacio de~Barros and G~Oas.
 Negative probabilities and counter-factual reasoning in quantum cognition.
  Physica Scripta, 2014(T163):014008, 2014.

\bibitem{AcacioNegative2}
G~Oas, J~Acacio de~Barros, and C~Carvalhaes.
 Exploring non-signalling polytopes with negative probability.
  Physica Scripta, 2014(T163):014034, 2014.

\bibitem{Holik-Plastino-ReviewNOVA-2014}
Federico Holik and Angel Plastino.
 Quantum mechanics: A new turn in probability theory.
 In Zoheir Ezziane, editor,  Contemporary Research in Quantum Systems, Physics Research and Technology.

\bibitem{Holik-Plastino-Saenz-ReviewHindawi}
Federico Holik, Cesar Massri, Manuel S\'{a}enz, and Angel Plastino. Generalized probabilities in statistical theories. arXiv:1406.0913 [stat.OT], 2014.

\bibitem{Barnum-CloningAndNoBroadcasting}
Howard Barnum, Jonathan Barrett, Matthew Leifer, and Alexander Wilce. Cloning and Broadcasting in Generic Probabilistic Theories. 2006.

\bibitem{HolevoBook-2011}
Alexander Holevo.
  Probabilistic and statistical aspects of quantum theory,
  volume~1 of  Quaderni. Monographs.
 Edizioni della Normale, Pisa, second edition, 2011.
 With a foreword from the second Russian edition by K. A. Valiev.

\bibitem{BubQLandQI}
Jeffrey Bub.
 Quantum computaton from a quantum logical perspective.
  Quantum Information And Computation, 7(4):281--296, May 2007.

\bibitem{GuhneCabello-2010}
Otfried G\"uhne, Matthias Kleinmann, Ad\'an Cabello, Jan-Ake Larsson,
  Gerhard Kirchmair, Florian Z\"ahringer, Rene Gerritsma, and Christian~F.
  Roos.
 Compatibility and noncontextuality for sequential measurements.
  Phys. Rev. A, 81:022121, Feb 2010.

\bibitem{CabelloProposalFor-2010}
Ad\'an Cabello.
 Proposal for revealing quantum nonlocality via local contextuality.
  Phys. Rev. Lett., 104:220401, Jun 2010.

\bibitem{CoxReview}
Kevin S Van Horn. Constructing a logic of plausible inference: a guide to cox's theorem. International Journal of Approximate Reasoning, 34(1):3--24, 2003.

\bibitem{CoxCriticas}
Stefan Arnborg and Gunnar Sjodin.
 On the foundations of bayesianism.
 In  AIP Conference Proceedings, pages 61--71. IOP INSTITUTE OF PHYSICS PUBLISHING LTD, 2001.

\bibitem{feynman1951}
Richard~P. Feynman.
  The Concept of Probability in Quantum Mechanics.
 University of California Press, Berkeley, Calif., 1951.

\bibitem{Holik-Plastino-PRA-2011}
F.~Holik and A.~Plastino.
 Convex polytopes and quantum separability.
  Phys. Rev. A, 84:062327, Dec 2011.

\bibitem{Holik-Zuberman-2013}
Federico Holik, César Massri, A.~Plastino, and Leandro Zuberman.
 On the lattice structure of probability spaces in quantum mechanics.
  International Journal of Theoretical Physics, 52(6):1836--1876, 2013.

\bibitem{Gleason1975}
AndrewM. Gleason.
 Measures on the closed subspaces of a hilbert space.
 In C.A. Hooker, editor,  The Logico-Algebraic Approach to Quantum Mechanics, volume~5a of  The University of Western Ontario Series in Philosophy of Science, pages 123--133. Springer Netherlands, 1975.

\bibitem{PeresBook}
Asher Peres.
  QuantumTheory: Concepts And Methods, volume~72 of  Fundamental Theories of Physics.

\bibitem{CavesFuchs-BayesianCoherence}
Christopher~A. Fuchs and R\"udiger Schack.
 Quantum-bayesian coherence.
  Rev. Mod. Phys., 85:1693--1715, Dec 2013.

\bibitem{Birkhoff-LatticeTheoryBook}
Garrett Birkhoff.
  Lattice theory, volume~25 of  American Mathematical Society
  Colloquium Publications.
 American Mathematical Society, Providence, R.I., third edition, 1979.

\bibitem{navara1991pasting}
Mirko Navara and Vladimir Rogalewicz.
 The pasting constructions for orthomodular posets.
  Mathematische Nachrichten, 154(1):157--168, 1991.

\bibitem{BeltramettiCassinelliBook}
Enrico~G. Beltrametti and Gianni Cassinelli.
  The logic of quantum mechanics, volume~15 of  Encyclopedia
  of Mathematics and its Applications.
 Addison-Wesley Publishing Co., Reading, Mass., 1981.
 With a foreword by Peter A. Carruthers.

\bibitem{BvN}
Garrett Birkhoff and John von Neumann.
 The logic of quantum mechanics.
  Ann. of Math. (2), 37(4):823--843, 1936.

\end{thebibliography}
\end{document}